\documentclass[twocolumn]{aastex63}
\usepackage{color}
\definecolor{mygreen}{rgb}{0.19,0.55,0.11}

\received{\today}
\revised{}
\accepted{}

\shorttitle{A MESA Model of a Double Neutron Star Progenitor}
\shortauthors{Jiang~et~al.}

\begin{document}
\title{Novel Model of an Ultra-stripped Supernova Progenitor of a Double Neutron Star}

\email{chenwc@pku.edu.cn}

\author{Long Jiang}
  \affil{School of Science, Qingdao University of Technology, Qingdao 266525, China}
  \affil{Department of Physics and Astronomy, Aarhus University, Ny Munkegade 120, DK-8000 Aarhus C, Denmark}
  \affil{School of Physics and Electrical Information, Shangqiu Normal University, Shangqiu 476000, China}
\author{Thomas M. Tauris}
  \affil{Department of Physics and Astronomy, Aarhus University, Ny Munkegade 120, DK-8000 Aarhus C, Denmark}
\author{Wen-Cong Chen}
  \affil{School of Science, Qingdao University of Technology, Qingdao 266525, China}
  \affil{School of Physics and Electrical Information, Shangqiu Normal University, Shangqiu 476000, China}
\author{Jim Fuller}
  \affil{TAPIR, Mailcode 350-17, California Institute of Technology, Pasadena, CA 91125, USA}

\begin{abstract}
Recent discoveries of double neutron star (DNS) mergers and ultra-stripped supernovae (SNe) raise the questions of their origin and connection. We present the first 1D~model of a DNS progenitor system which is calculated self-consistently until an ultra-stripped iron core collapse. We apply the \texttt{MESA} code starting from a post-common envelope binary consisting of a $1.35\;M_\odot$ NS and a $3.20\;M_\odot$ zero-age main-sequence helium star and continue the modelling via Case~BB Roche-lobe overflow until the infall velocity of the collapsing iron core exceeds $1000\;{\rm km\,s}^{-1}$. The exploding star has a total mass of $\sim 1.90\;M_\odot$, consisting of a $\sim 0.29\;M_\odot$ He-rich envelope embedding a CO core of $\sim 1.61\;M_\odot$ and an iron-rich core of $\sim 1.50\;M_\odot$. The resulting second-born NS has an estimated mass of $\sim 1.44\;M_\odot$ and we discuss the fate of the post-SN system, as well as the mild recycling of the first-born NS.
Depending on the initial conditions, this family of systems is anticipated to reproduce the DNS mergers detected by the LIGO-network.
\end{abstract}


\section{Introduction}
A new era of astrophysics began six years ago when the advanced Laser Interferometer Gravitational-Wave Observatory (LIGO) detected the first high-frequency gravitational wave (GW) signals from the double black hole (BH) merger event GW150914 \citep{abbo16}. Subsequently, another $\sim 50$ double BH mergers have been detected \citep{LVC21}, as well as a couple of double neutron star (NS) \citep{abbo17a,aaa+20} and mixed BH/NS mergers \citep{LVK21}.
The remarkable double NS (DNS) merger event GW170817 was not only detected in GWs, but also in the entire electromagnetic spectrum from gamma-rays to radio waves, marking the beginning of a new era of multi-messenger astronomy \citep{abbo17b}.

DNS systems play a vital role in a broad range of aspects of modern astrophysics.
First, they are important probes of exotic binary stellar evolution and provide key information of their past evolutionary history via: survival of multiple stages of mass transfer in X-ray binaries, one or more common envelope (CE) phases, as well as two supernova (SN) explosions \citep{vdH76,dpsv02,belc02,voss03,pods04,taur17}.
Second, high-precision radio pulsar timing of several DNSs can measure relativistic post-Keplerian parameters and thereby provide important tests for theories of gravity \citep[e.g.][]{ksm+06} as well as accurately determine the masses of the pulsar and its companion.
Third, GW merger events of DNSs involve baryonic matter (unlike double BH mergers). Hence, besides producing a short $\gamma$-ray burst \citep{abbo17b} these events lead to production of heavy chemical elements (beyond iron), through r-process nucleosynthesis, which decay and power an optical transient \citep[i.e. a {\em kilonova}, e.g.][]{scj+17,met19}.
Finally, DNS mergers may be related to FRBs via production of stable magnetars or pre-merger magnetospheric interactions \citep[e.g.][]{mbm19,zhan20}.

The standard formation scenario of DNS systems is as follows \citep{bhat91,taur17,tv22}: a massive binary with a pair of OB-stars first experiences an episode of stable mass transfer via Roche-lobe overflow (RLO) from the primary star to the secondary star, before the former (initially most massive) star explodes in a Type~Ib/c SN \citep{yoon10}.
The system is subsequently observable as a high-mass X-ray binary (HMXB) with a NS orbiting an OB-star \citep{lvv00}.
Once the secondary star evolves to become a giant and initiates RLO towards the accreting NS, the system becomes dynamically unstable due to the large mass ratio between the stellar components and it enters a phase of common-envelope (CE) evolution \citep{pacz76,ijc+13}. After in-spiral of the NS and successful ejection of the hydrogen-rich CE, the NS and the naked He~star (the remaining core of the secondary giant star) appear in a tight orbit. Subsequently, an additional stage of stable RLO (Case~BB RLO) often occurs from the expanding He~star to the NS \citep{habe86,dpsv02,taur15}.
In this process, the NS is mildly recycled to a short spin period of order $10-100\;{\rm ms}$ while the extreme stripping process of the He~star subsequently leads to a so-called ultra-stripped SN explosion \citep{taur13,taur15,suwa15,mori17,mgh+18,mth+19}.

In case the system survives this second SN explosion, an eccentric DNS is formed which will merge within a Hubble time if the post-SN orbital period of the DNS system is short enough ($\la 0.7\;{\rm d}$) or its eccentricity is large.
For fairly tight pre-SN systems, and given that the NS kick resulting from an ultra-stripped SN is often small \citep[although not always, depending on the NS mass,][]{taur17,mth+19} the post-SN system is indeed expected to remain bound in most cases.

\cite{taur13,taur15} proposed that all DNS mergers formed in the Galactic disk should arise from an ultra-stripped SN explosion of the secondary star. In the former paper, a first detailed binary star calculation was presented in which a pre-SN stellar mass is barely above the Chandrasekhar limit ($\sim 1.4\;M_\odot$), resulting in the ejection of only $\sim 0.05-0.20\;M_\odot$ of material from the explosion --- an order of magnitude smaller than that of typical Type~Ib/Ic~SNe \citep{pdc+10,drou13}. They also computed the light curve of such an ultra-stripped SN of Type~Ic and demonstrated that it is expected to be extremely fast and faint.
The follow-up and more detailed work on the spectral properties by \citet{mori17}, suggested that ultra-stripped SNe synthesize of order $\sim 0.01\;M_\odot$ of radioactive $^{56}$Ni, and their typical peak luminosities were estimated to be around $10^{42}\;{\rm erg\,s}^{-1}$, or $-16\;{\rm mag}$.
The light curve evolution becomes faster with smaller ejecta mass \citep[because of the smaller diffusion timescale which is roughly proportional to $(\Delta M_{\rm ejecta}^3/E_{\rm SN})^{1/4}$, where $\Delta M_{\rm ejecta}$ is the ejecta mass and $E_{\rm SN}$ is the kinetic energy of the SN, e.g.][]{arn82}.
Based on such rapidly decaying light curves, several potential ultra-stripped SN candidates have already been found \citep{drou13,mori17,de18}.

Employing the BEC stellar evolution code \citep[e.g.][]{yoon10}, \citet{taur13} were only able to modeled the ultra-stripped progenitor star (i.e. the naked He~core evolving from a NS+He~star system) until the ignition of off-center oxygen burning. Using the Modules for Experiments in Stellar Astrophysics \citep[\texttt{MESA},][]{paxt11}, \cite{mori17} successfully evolved a He~star from its zero-age main-sequence (ZAMS) stage until the formation of an iron core that collapsed. However, their simulations were based on single-star evolution that mimicked binary evolution by treating mass-loss in a simple way, following the mass exchange in the binary model of \citet{taur13}. Therefore, up to now a self-consistent modelling of the evolution of a NS+He~star system until the core collapse of such an ultra-stripped He~star has been missing in the literature. Here, we present such a calculation for the first time.

\section{Method}
To obtain a detailed evolutionary model of a DNS progenitor system, we simulated the evolution of a NS+He-star binary using the \texttt{MESA} code module \texttt{MESAbinary} \citep[r12778,][and references therein]{paxt19}.
The starting point of the simulation is a binary system containing a NS with a initial mass of $M_{\rm NS,i}=1.35\;M_\odot$ and a He-ZAMS companion star (with a chemical composition of $Y=0.98$ and $Z=0.02$) with an initial mass of $M_{\rm He,i}=3.20\;M_\odot$ in a circular orbit with initial orbital period of $P_{\rm orb,i}=1.0\;{\rm d}$.
We adopted a mixing-length parameter, $\alpha=l/H_{\rm p}=1.5$ \citep{lang91}, where $l$, and $H_{\rm p}$ denote the mixing length and the local pressure scale height, respectively. \citep[Applying e.g. $\alpha=2$ only increases the final metal core mass by $<2\%$,][]{taur15}.
For the wind mass-loss rate of the He~star, we used the ``Dutch'' prescription \citep{gleb09} with a scaling factor of 1.0, and Type2 opacities were applied for $Z=0.02$.
For the nuclear reactions, we adopted from the start the large \texttt{MESA} nuclear network \texttt{mesa151.net}, which includes 151 nuclei, following the modelling of \citet{mori17}.

To compute the mass-transfer rate from the He~star ($|\dot{M}_{\rm He}|$), we adopted the ``kolb'' mass-transfer scheme based on optical thick overflow \citep{kolb90}.
During the mass transfer, we adopted a fixed Eddington accretion rate of $\dot{M}_{\rm Edd}=4.0\times10^{-8}\;M_\odot\,{\rm yr}^{-1}$ as an upper limit for the accumulation of He-rich material onto a NS (notice, for Case~BB RLO: $|\dot{M}_{\rm He}|\gg \dot{M}_{\rm Edd}$).
The orbital evolution during RLO was solved using the "isotropic re-emission" model \citep{bhat91,tv22} which assumes that the fraction of accreted material arriving at a rate in excess of $\dot{M}_{\rm Edd}$ is re-emitted as an isotropic fast wind with the specific orbital angular momentum of the NS. Orbital angular momentum loss due to GWs was included, but is negligible here.

The numerical modelling of a binary low-mass He~star until iron core collapse is challenging.
Therefore, we divided the evolution of the donor star into three stages\footnote{Our inlists are available at:\\ \url{https://zenodo.org/record/5516837\#.YUiMDMi_OyJ}.\\
Note, the only difference in inlist between stages 1 and 2 is the value of: \texttt{op\_split\_burn\_min\_T}.}:
\begin{itemize}
\item [i)] The first $N=2000$ models until oxygen burning is ignited (slightly off-center; $\sim 7\;{\rm yr}$ prior to core collapse) and a Si~core begins to form, $f_{\rm Si}>0.02$. (Without interruption, this first stage may otherwise proceed until $N\sim 2400$ and $f_{\rm Si}\sim 0.3$. $|\dot{M}_{\rm He}|$ is unaffected until a few months prior to SN.)
\item [ii)] The second stage is continued from the previous model until the central temperature, $T_{\rm c}>10^{9.9}\;{\rm K}$ ($N\sim 4300$, of order 1~min prior to core collapse).
\item [iii)] The third stage continues the evolution until iron core collapse (i.e. when the infall velocity reaches $v_{\rm in}\sim 1000\;{\rm km\,s}^{-1}$ in model $N=4563$).
\end{itemize}

\section{Results}

\begin{figure}
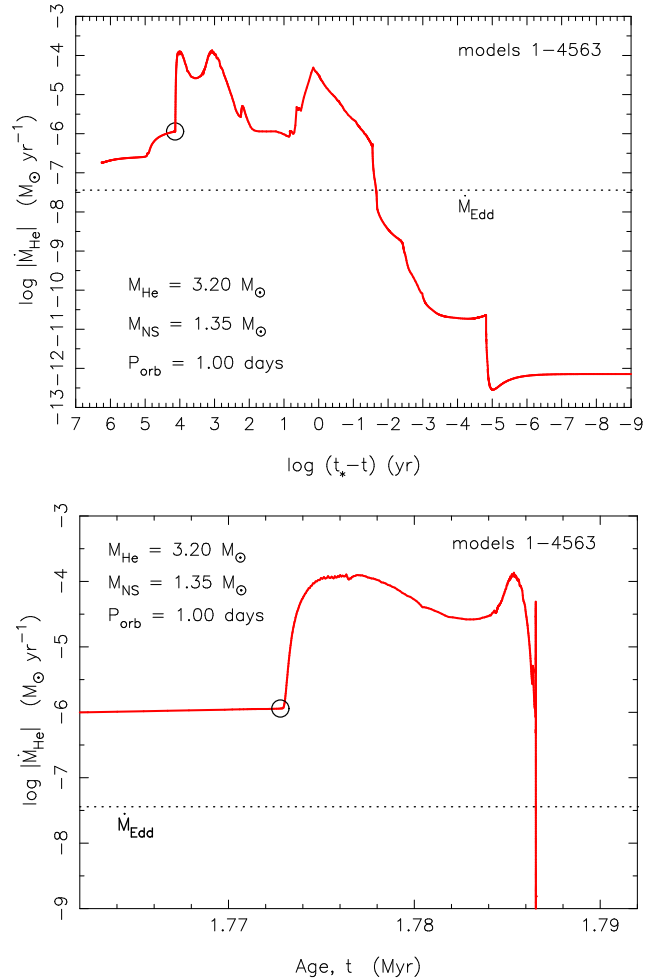

\centering
 \includegraphics[width=0.35\textwidth, angle=-90]{f1a.eps}\vspace{0.30cm}
 \includegraphics[width=0.35\textwidth, angle=-90]{f1b.eps}
\caption{Mass-loss rate of He~star donor versus remaining time until core collapse (top) or stellar age (bottom) of the simulated NS+He~star system. The total simulated time from He-ZAMS to iron core collapse is $t_{\ast} = 1.786536\;{\rm Myr}$. The horizontal dashed line represents our assumed Eddington accretion rate for helium ($\dot{M}_{\rm Edd}=4.0\times10^{-8}\;M_\odot\,{\rm yr}^{-1}$). The open circles denote the onset of Case~BB RLO.}
\label{fig:mass-loss-rate}
\end{figure}

\begin{figure}
\centering
\vspace*{-0.30cm}\hspace*{-0.6cm}\includegraphics[width=1.2\linewidth,trim={0 0 0 0},clip]{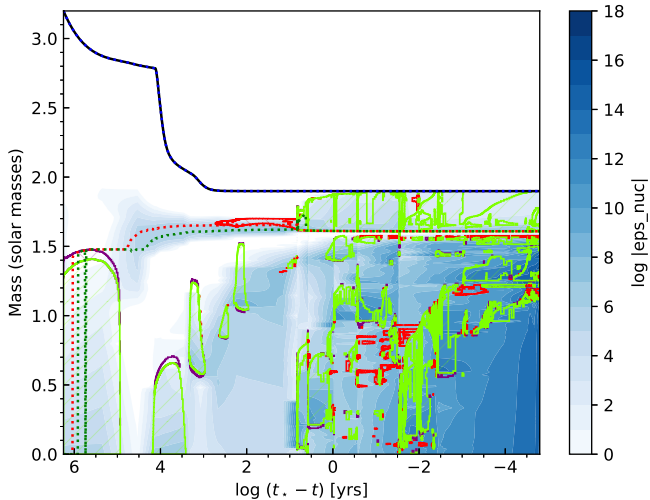}
\caption{Kippenhahn diagram of the He~star with $M_{\rm He,i}=3.20\;M_\odot$ experiencing Case~BB RLO (onset at $\log (t_\ast -t)\simeq 4.1$) in Figures~\ref{fig:mass-loss-rate} and \ref{fig:temp-density-HR}. The plot shows cross-sections of the He~star in mass-coordinates from the center to the surface of the star, along the y-axis, and as a function of remaining stellar age until core collapse on the x-axis. The green hatched areas denote zones with convection; red color indicates semi-convection. The red (green) dotted line indicates the boundary of the carbon (oxygen) core. The intensity of the blue color indicates the nuclear energy production rate.}
\label{fig:kippenhahn}
\end{figure}

\begin{figure}
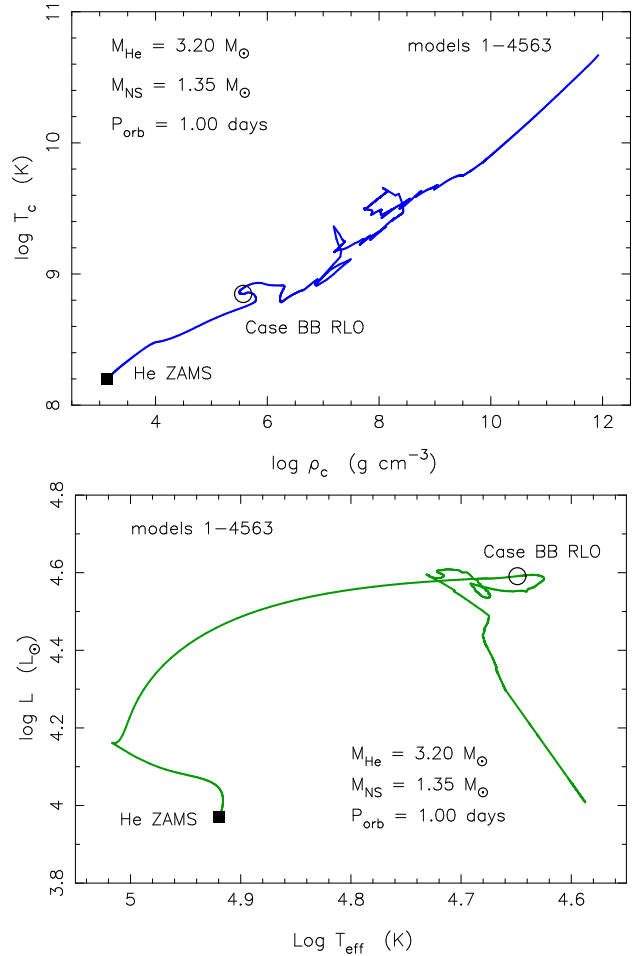

\centering
\includegraphics[width=0.35\textwidth, angle=-90]{f3.eps}
\includegraphics[width=0.35\textwidth, angle=-90]{f4.eps}
\caption{Evolutionary track of the He~star in the central temperature versus central density diagram (top) and in the HR~diagram (bottom). The solid square and open circle denote the He-ZAMS and onset of Case~BB RLO, respectively.}
\label{fig:temp-density-HR}
\end{figure}

Using the \texttt{MESA} code, we simulated the evolution of a binary system consisting of a $M_{\rm NS,i}=1.35\;M_{\odot}$ NS and a $M_{\rm He,i}=3.20\;M_\odot$ ZAMS He~star with an initial orbital period of $P_{\rm orb,i}=1.0\;{\rm d}$.
Figure~\ref{fig:mass-loss-rate} shows a plot of the evolution of the He~star and the mass-loss rate, $|\dot{M}_{\rm He}|$ versus remaining time until core collapse, $\log (t_\ast -t)$ (top) or stellar age, $t$ (bottom). The total simulated stellar age from He-ZAMS to iron core collapse is $t_{\ast} = 1.786536\;{\rm Myr}$ and Case~BB RLO is initiated at $t\simeq 1.773\;{\rm Myr}$.
Before the onset of the RLO, the rate of stellar wind mass loss from the He~star gradually increases from $10^{-7}-10^{-6}\;M_\odot\,{\rm yr}^{-1}$, thereby increasing $P_{\rm orb}\simeq 1.21\;{\rm d}$. Once mass transfer via RLO is initiated, $|\dot{M}_{\rm He}|\simeq 10^{-4}\;M_\odot\,{\rm yr}^{-1}$ and thus $|\dot{M}_{\rm He}|\gg \dot{M}_{\rm Edd}$.

Figure~\ref{fig:kippenhahn} shows the Kippenhahn diagram of the ultra-stripped He~star donor, from the He-ZAMS to iron core collapse. As a result of the strong stellar wind mentioned above, the He~star decreases its mass from 3.20 to $2.79\;M_\odot$ by the time it fills its Roche lobe at $\log (t_\ast -t)\simeq 4.1$ and initiates Case~BB RLO \citep[technically speaking, this is Case~BC RLO as mass transfer is not initiated before the onset of core carbon burning,][]{taur15}. At that time, i.e. $\sim 13\,500\;{\rm yr}$ prior to iron core collapse, the central abundances of carbon and oxygen are 0.32 and 0.60, respectively.
Oxygen burning ignites slightly off-center at about 5~yr prior to iron core collapse. However, as opposed to the BEC models of \citet{taur15}, our simulation can continue further and ignite silicon burning at $\log (t_\ast -t)\simeq -1.5$, or about 11~d prior to the iron core collapse.

Figure~\ref{fig:temp-density-HR} (top) presents the evolution of central temperature ($T_{\rm c}$) versus central mass-density ($\rho_{\rm c}$) of the ultra-stripped SN progenitor. When $\log T_{\rm c}=9.07$ oxygen is ignited, and at $\log T_{\rm c}=9.54$ silicon begins to burn and subsequently an iron core forms.
At the moment of the core collapse, the central density and temperature are $\log\rho_{\rm c}=11.88$ and $\log T_{\rm c}=10.64$, respectively.
These values are higher than those obtained by \citet{mori17}. This difference is most likely to originate from different masses of the exploding stars.

Figure~\ref{fig:temp-density-HR} (bottom) displays the evolutionary track of the He~star
donor in the HR~diagram. When the stellar age reaches $t\simeq 1.70\;{\rm Myr}$ (corresponding to a luminosity of $\log (L/L_\odot)\simeq 4.16$), central helium is exhausted, and the He~star climbs the giant branch and expands while undergoing helium shell burning. Case~BB RLO is initiated at $\log (L/L_\odot)\simeq 4.59$ and subsequently the luminosity of the He~star decreases by a factor of $\sim 4$ prior to core collapse.

The final mass of the ultra-stripped SN progenitor\footnote{We apply the term ``ultra-stripped SNe'' in a slightly broad context to include all exploding stars in tight binaries (also some with He envelopes $>0.2\;M_\odot$) producing potential DNS mergers.} at the moment of core collapse (SN explosion) is about $1.90\;M_\odot$, and that of the NS is $\sim 1.35\;M_\odot$ ($1.35051\;M_\odot$, see below). The final orbital period of our simulated NS+He~star systems is 0.872~d.
The exploding star has a CO~core of $\sim 1.61\;M_\odot$ and an iron-rich core of $\sim 1.50\;M_\odot$.
Hence, it has a He-rich envelope of $0.29\;M_\odot$ with mass fractions of 0.84 (He), 0.11 (C) and 0.05 (O). The total amount of helium ejected in the SN is thus $\sim 0.24\;M_\odot$, which exceeds the minimum mass threshold of $\sim 0.06\;M_\odot$ that makes helium lines visible in optical/IR spectra \citep{hach12}; see also \citet{dyal20} for discussions. Therefore, the SN resulting from our simulations should be observable as a Type~Ib rather than Type~Ic.

During the Case~BB X-ray phase, with a duration of $\Delta t_X=13\,500\;{\rm yr}$, the NS accreted a total of $\Delta M_{\rm NS}=5.1\times 10^{-4}\;M_\odot$. The modest amount of accreted material will result in a mildly recycled NS with an expected spin period of at least 102~ms \citep[see Eq.~14 in][]{taur12}. Although not fully recycled to a millisecond pulsar, the B-field of this first-born NS is still expected to be sufficiently reduced to about $B\simeq 3-10\times 10^9\;{\rm G}$ as a result of the mass accumulated \citep{taur17}, which results\footnote{Here assuming $B\simeq 10^{19}\;{\rm G}\;\sqrt{P\dot{P}}$ \citep{taur17}.} in a spin-down age, $\tau \equiv P/(2\dot{P})$ of order $0.16-1.8\;{\rm Gyr}$. Hence, the DNS system formed after the second SN is expected to have an active radio lifetime of the same order.

\begin{figure}
\centering
\vspace*{-0.8cm}\hspace*{-0.8cm}\includegraphics[scale=0.53]{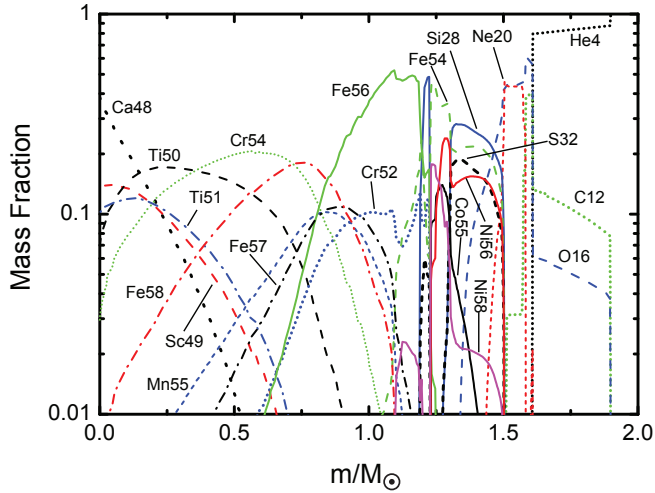}\vspace{-4.0cm}
\caption{Mass fractions of chemical abundances as a function of the mass coordinate of the ultra-stripped iron core collapse SN progenitor. The data comes from our last simulated model ($N=4563$) at $t\simeq t_\ast = 1.786536\;{\rm Myr}$. The mass-density profile is shown in Figure~\ref{fig:density}.}
\label{fig:abundances}
\end{figure}

Figure~\ref{fig:abundances} illustrates the final chemical structure of the $\sim 1.90\;M_\odot$ ultra-stripped He~star in our last model ($N=4563$) at the moment of core collapse. Inside the boundary of the $\sim 1.50\;M_\odot$ iron-rich (Fe/Si) core, besides $^{56}$Fe, the innermost $1.20\;M_\odot$ has produced many neutron-rich elements, including: $^{48}$Ca, $^{50}$Ti, $^{51}$Ti, $^{54}$Cr $^{58}$Fe, etc.
The density structure of the last model is plotted in Figure~\ref{fig:density}. Within a radius of $\sim 10^{-4}\;R_\odot$, the mass density reaches $\sim 10^{12}\;{\rm g\,cm}^{-3}$, which is about two orders of magnitude higher than that obtained by \citet{mori17}.

\section{Discussion}
Among ultra-stripped SNe which have been discovered so far \citep[e.g. SN~2005ek, SN~2010X and iPTF~14gqr,][]{drou13,mori17,de18} the latter event indicates, from early photometry and spectroscopy, evidence of shock cooling of an extended helium-rich envelope ($M_{\rm env}\sim 0.01\;M_\odot$, $R_{\rm env}\ga 500\;R_\odot$), likely ejected in an intense pre-explosion mass-loss episode of the progenitor $\sim 1$~to~3~weeks prior to the explosion --- about at the same time as silicon flashes occur in the terminal evolution of low-mass metal cores  \citep{wh15}. Further investigations are needed to explore both this possibility of explosive ignition of degenerate elements in the central regions or wave-driven outbursts \citep{qs12,fr18}.
We did not see such an eruption in our model, but they may occur in slightly lower mass stars that become more degenerate in their cores, and further investigation is needed.

\begin{figure}
\centering
 \includegraphics[width=0.35\textwidth, angle=-90]{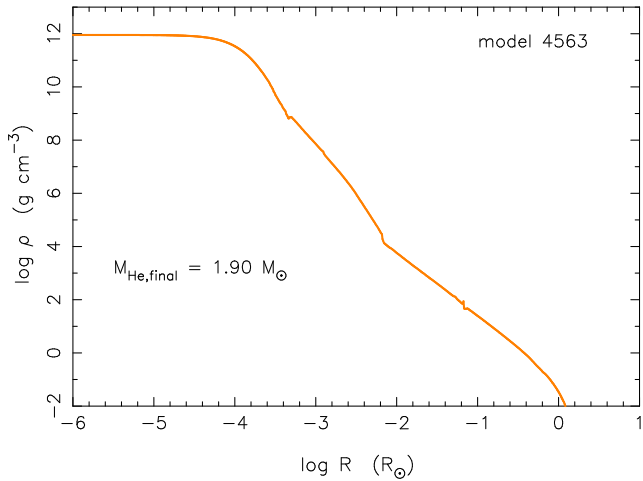}
\caption{Mass-density profile of the simulated ultra-stripped SN progenitor at the moment of iron core collapse. The final He~star has a total mass of $\sim 1.90\;M_\odot$.}
\label{fig:density}
\end{figure}

It is noteworthy that the shown simulated mass-transfer rate just prior to the SN explosion reaches a very low value of $<10^{-10}\;M_\odot\,{\rm yr}^{-1}$ (even reaching $\sim 10^{-12}\;M_\odot\,{\rm yr}^{-1}$ at the very moment of the SN explosion).
A caveat in our modelling, however, is that the final mass-transfer rate is somewhat sensitive to our choice of the parameter \texttt{op\_split\_burn\_min\_T}, which allows MESA to use a higher order solver to help it converge in regions with high temperature, and rather strongly dependent on our model number for the transition from stage i to ii.
Hence, the calculated value of $|\dot{M}_{\rm He}|$ during the last few months prior to core collapse is uncertain. Importantly, this parameter has little influence on the final mass-density profile shown in Fig.~\ref{fig:density}.

Although, some material ejected due to intense RLO is still present in the near-environment of the SN (as $|\dot{M}_{\rm He}|\sim 10^{-5}\;M_\odot\,{\rm yr}^{-1}$ up to less than 1~yr before the SN, Figure~\ref{fig:mass-loss-rate}), it is still a relatively insignificant amount.
It is possible that some ultra-stripped SN progenitors will have more intense pre-SN mass loss due to RLO, as the mass-transfer rate in these stars is determined by the helium shell-burning luminosity, which can in turn be affected by the precise sequence of the carbon, neon, and oxygen burning in the core, as well as the possibility of vigorous helium shell flashes near the surface layer \citep{taur13,taur15}.

Future work should include a systematic investigation of NS+He~star systems of different masses and orbital periods as these parameters have been demonstrated \citep[][although not calculated beyond oxygen burning]{taur15} to yield different outcomes in terms of remaining envelope mass, efficiency of recycling and final mass-transfer rate (or even complete decoupling from the Roche lobe). Early attempts from our side indicate that numerical difficulties may arise even when applying slightly different initial conditions.
A full discussion is beyond the scope of this Letter.

The X-ray lifetime ($\Delta t_X \simeq 13\,500\;{\rm yr}$) of this NS+He~star system is shorter than that of 40\,000 yr for the model investigated in \citet{taur13} and \citet{mori17}.
The discrepancy arises from using different initial He~star masses and orbital periods. In our work presented here, we started out with $M_{\rm He,i}=3.20\;M_\odot$ and $P_{\rm orb,i}=1.0\;{\rm d}$, compared to $M_{\rm He,i}=2.90\;M_\odot$ in a very tight orbit with $P_{\rm orb,i}=0.1\;{\rm d}$ in the latter work. \citet{taur15} demonstrated that there exists an anti-correlation between the mass-transfer timescale, $\Delta t_X$ and initial orbital period, $P_{\rm orb,i}$. A long initial orbital period causes the He~star to be more evolved by the time it initiates Case~BB RLO, which reduces its remaining lifetime until core collapse, and thereby reduces the amount of material transferred to the accreting NS.
This results in a post-SN correlation between the observed spin period of the recycled (first-born) NS and orbital period of the DNS system \citep{taur17}.

An important question is whether or not the post-SN DNS system will merge within a Hubble time.
If we assume that the CO (metal) core boundary at $1.61\;M_\odot$ is a rough estimate for the baryonic mass cut of the second-born NS, $M_{\rm NS}^{\rm baryon}$, the gravitational mass of this NS is $M_{\rm NS,2}=M_{\rm NS}^{\rm baryon}-\Delta M_{\rm NS}^{\rm bind}$, where the release of gravitational binding energy of the newborn NS is given by \citep{latt89}:
\begin{equation}\label{eq:Ebind}
    \Delta M_{\rm NS}^{\rm bind}\simeq 0.084\;M_\odot \;(M_{\rm NS,2}/M_\odot)^2
\end{equation}
Thus, we find an expected (gravitational) mass of the newborn secondary NS to be $M_{\rm NS,2}\simeq 1.44\;M_\odot$. This is a relatively high value, but the total mass of the DNS system is $2.79\;M_\odot$, which is not an unusual among the observed Galactic DNS systems \citep[e.g.][]{taur17}. If naively solving the equations of \citet{pet64} for the merger time due to emission of GWs of a circular DNS system using a pair of NSs with masses $M_{\rm NS,1}=1.35\;M_\odot$ and $M_{\rm NS,2}=1.44\;M_\odot$, and assuming $P_{\rm orb}=0.872\;{\rm d}$ to remain unchanged during the SN, the resulting, merger time is $\tau_{\rm GW}\sim 23.7\;{\rm Gyr}$. However, the possibility of a tighter post-SN orbit and, in particular, an orbital eccentricity, $e$ resulting from the SN can significantly decrease $\tau_{\rm GW}$. In fact, $\tau_{\rm GW}$ will decrease by more than a factor of 10 if $e>0.694$ for the same $P_{\rm orb}$.

Due to an explosion asymmetry, the newborn NS may receive a significant momentum kick during the SN \citep{jan12}.
For our pre-SN system, we have an orbital separation of $a=5.69\;R_\odot$, a total mass of $M=3.25\;M_\odot$ and a relative velocity between the stellar components of $v_{\rm rel}=330\;{\rm km\,s}^{-1}$. An amount of $\Delta M=0.46\;M_\odot$ is ejected during the SN and thus we find at least 23\% probability for a smaller post-SN orbit for all kick magnitudes between $50<w<300\;{\rm km\,s}^{-1}$ \citep[using eq.13 of][]{taur17}. Monte Carlo simulations reveal a $\sim 30\%$ probability of a post-SN merger within a Hubble time if $w\simeq 200-250\;{\rm km\,s}^{-1}$.
Thus, we see that for a range of reasonable kicks, there is a relatively large probability that a DNS merger is produced.
For the kicks in the second SN in DNS systems, \citet{taur17} concluded from their analysis of known DNS parameters that $0\la w\la300\;{\rm km \,s}^{-1}$, where the larger kicks are apparently imparted onto the somewhat more massive NSs (e.g. PSR~B1913+16, or the system presented here). In addition, the SN imparts an eccentricity on the post-SN orbit which can take any value $0\la e < 1$ for the surviving orbits, depending on kick direction. We thus conclude that our simulated system is a potential example of a progenitor system of a DNS merger event.

\section{Conclusions}
We calculated the first detailed and self-consistent 1D stellar evolution model of a DNS progenitor system until the ultra-stripped iron core collapse in the second SN event. We applied the \texttt{MESA} code on a post-CE binary containing a $3.20\;M_\odot$ He-ZAMS donor star and a $1.35\;M_\odot$ NS with an initial orbital period of $1.0\;{\rm d}$. The final pre-SN system is an ultra-stripped He~star of $1.90\;M_\odot$ (with a CO core of $1.61\;M_\odot$ and an iron core of $1.50\;M_\odot$) which, at the moment of core collapse (modelled until the infall velocity exceeds $1000\;{\rm km\,s}^{-1}$), orbits a mildly recycled pulsar with a spin period of $\sim 100\;{\rm ms}$ and $P_{\rm orb}=0.872\;{\rm d}$. The ultra-stripped SN is in this case expected to leave a second-born NS with a mass of $\sim 1.44\;M_\odot$.
The \texttt{MESA} modelling presented here opens for a new avenue of systematic investigations of similar DNS progenitors. In the particular example demonstrated here, the post-SN DNS system has a total mass of $2.79\;M_\odot$. In comparison, GW170817 has a total mass of $M=2.74^{+0.04}_{-0.01}\;M_\odot$ \citep{abbo17a}.
To produce a massive DNS merger like GW190524 \citep[$M\sim 3.4\;M_\odot$,][]{aaa+20,kru20} would require e.g. a more massive first-born NS, similar to the $\sim 1.8-2.0\;M_\odot$ NSs in the HMXBs Vela~X-1 \citep{rom+11} and IGR~J17252$-$3616 \citep{mwb12}.
The timescale for GW damping of a DNS system until it merges depends strongly on the imparted kick magnitude and its direction. During the last $\sim 10-50\;{\rm Myr}$ of in-spiral, a similar Galactic source would be detectable by the space-borne low-frequency GW detectors LISA, Taiji, and TianQin.

\acknowledgments {We are grateful to the anonymous reviewer for constructive comments improving our manuscript. We thank Takashi Moriya for useful discussions. This work was partly supported by the National Natural Science Foundation of China (under grant Nos. 11573016, 11803018, and 11733009) and the CAS ``Light of West China''
Program (grant No. 2018-XBQNXZ-B-022).}

\end{document}